\documentclass[twocolumn,prb,showpacs]{revtex4}
\AtBeginDocument{
\heavyrulewidth=.08em
\lightrulewidth=.05em
\cmidrulewidth=.03em
\belowrulesep=.65ex
\belowbottomsep=0pt
\aboverulesep=.4ex
\abovetopsep=0pt
\cmidrulesep=\doublerulesep
\cmidrulekern=.5em
\defaultaddspace=.5em
}

\usepackage{amsmath,amssymb}
\usepackage{graphicx}
\usepackage[usenames,dvipsnames]{xcolor}
\usepackage{amsthm}
\usepackage{booktabs}
\usepackage{hyperref}

\definecolor{myred}{HTML}{FF4136}

\hypersetup{colorlinks=true,citecolor=Orange,linkcolor=Red,urlcolor=Black}

\begin{document}

\title{Spectral Functions with the Density Matrix Renormalization Group:\\
Krylov-space Approach for Correction Vectors}

\author{A. Nocera  and G. Alvarez}
\affiliation{Computer Science \& Mathematics %
Division and Center for Nanophase Materials Sciences, Oak Ridge National Laboratory, %
 \mbox{Oak Ridge, Tennessee 37831}, USA}

\begin{abstract}
Frequency-dependent correlations, such as the spectral function and the dynamical structure
factor, help understand condensed matter experiments.
Within the density matrix renormalization group (DMRG) framework, 
an accurate method for calculating spectral functions directly 
in frequency is the correction-vector method. The correction-vector can be 
computed by solving a linear equation or by minimizing a functional. 
This paper proposes an alternative to calculate the correction vector: 
to use the Krylov-space approach.
This paper then studies the accuracy and performance of the Krylov-space approach, when
applied to the Heisenberg, the t-J, and the Hubbard models.
The cases studied indicate that Krylov-space approach can be more accurate and
efficient than conjugate gradient, and that the error of the former integrates best
when a Krylov-space decomposition is also used for ground state DMRG.
\end{abstract}


\maketitle
\newpage

\section{Introduction}

In the last two decades, several approaches have been proposed for the calculation of 
frequency dependent
correlations or spectral functions of low dimensional strongly correlated systems
within the density matrix renormalization group (DMRG) method.\cite{re:White1992,re:White1993}
 In 1995, Hallberg proposed a Lanczos-vector method based on a continued 
fraction expansion.\cite{re:Hallberg1995} 
Later, K\"{u}hner and White~\cite{re:Kuhner1999} 
showed that this method is suitable only for spectra consisting of few discrete peaks. 
The same paper introduced the correction vector method to calculate
spectral functions. Jeckelmann later developed a variational 
improvement to the correction-vector method---the dynamical 
DMRG.\cite{re:Jeckelmann2002}
It has since been used successfully
in several studies\cite{re:Jeckelmann2008} to calculate the dynamical properties, 
 directly in frequency. The main disadvantage of the correction-vector method is two-fold: (i) dynamical properties need to be computed in small intervals of 
frequency; (ii) the artificial broadening, $\eta$, which always needs to be introduced
at some point in the calculation, is set from the start, and to change it a new run needs to be carried out.
The $\eta$ broadening
can be viewed as convolution 
of the exact spectral function with a Lorentzian of the same width. 
Because one is interested in the exact 
spectrum, one needs to perform a deconvolution of the spectrum---an ill-defined operation. 
Many works have presented different and successful deconvolution procedures for correction-vector
DMRG spectra of one dimensional\cite{re:Gebhard2003,re:Ulbricht2010} and quantum impurity
 problems.\cite{re:Weichselbaum2009,
 re:Nishimoto2004,re:Raas2005,re:Raas2004} Recently, 
ref.~\onlinecite{re:Paech2014} has proposed a blind deconvolution algorithm, having
the advantage of reproducing quite well sharp singularities such as power-law band edges and
excitonic peaks, but the drawback of introducing artificial shoulder-like spectral features.

Within the DMRG framework, another method for calculating the 
dynamical properties consists of first computing the correlation functions using 
time-dependent 
DMRG,\cite{re:Vidal2004,re:Feiguin2004,re:Kollath2004} and then 
space-time Fourier transforming them into real frequency. Recently, 
ref.~\onlinecite{re:Holzner2011} presented a highly 
efficient and accurate adaptive method using 
Chebyshev polynomials in combination with matrix product states (MPS).
In order to compute time-dependent correlation functions over a long time interval, 
ref.~\onlinecite{re:White2008} has shown that one can apply an 
efficient prediction method.
In addition, ref.~\onlinecite{re:Wolf2015} has shown that linear prediction can be used
as a method to extrapolate Chebyshev or Fourier expansions of spectral functions. 
Within the MPS formulation of DMRG, 
other powerful methods have been 
developed.\cite{re:Dargel2011,re:Holzner2011,re:Dargel2012}
Ref.~\onlinecite{re:Dargel2011} and \onlinecite{re:Dargel2012} have proposed 
an improvement of the continued fraction 
expansion method of Hallberg in the MPS language.  

In this paper, we focus on the correction-vector 
DMRG method, and propose an alternative method to calculate the correction vector. 
In its traditional formulation, the imaginary part of the correction-vector, being directly
proportional to the spectral function, is the solution of an 
inhomogeneous set of linear equations. This set of equations can be solved explicitly with the 
conjugate-gradient method. The conjugate-gradient method's convergence error
is independent of the ground state DMRG error.
Dynamical DMRG\cite{re:Jeckelmann2002} recasts the calculation of the correction-vector 
as an elegant variational principle, defining a functional that one then minimizes. 
The main advantage of dynamical DMRG is the possibility of finding a solution with 
an error smaller than that of the traditional correction-vector method. 

We here propose an alternative approach: to calculate the correction vector
using a Krylov-space\cite{re:krylov31} decomposition instead of solving a linear 
equation. 
Krylov space decomposition approaches have found application in several fields of science and 
engineering, where the computation of functions of large sparse matrices is often the main problem to be solved. 
In physics, they have been used in several 
contexts\cite{re:Alvarez2011} as a technique to solve partial or ordinary differential equations, such as numerical general relativity and electrodynamics.\cite{re:Borner2008}

The correction vector is not then calculated with a \emph{separate} algorithm, be it conjugate gradient 
as originally proposed,  
or minimization of an auxiliary functional as in dynamical DMRG, but with the following algorithm instead.
At each DMRG step 
(i) tri-diagonalize and successively diagonalize the Hamiltonian of the problem in 
the current basis; 
(ii) calculate the correction-vector in the diagonal basis and rotate it back to 
the current basis.
The error profile is then the same as standard DMRG if
Krylov-space decomposition is used for the ground state computation, because
Krylov-space decomposition is needed at each step of the ground-state DMRG 
algorithm. Our proposal therefore integrates better with 
ground-state DMRG. Translated into MPS language, our method 
works on the same space spanned by the effective Hamiltonian where all but one
site of the tensor network (MPO-MPS product) has been contracted,
where MPO stands for matrix product operator.
Moreover, to avoid Lanczos ghost 
states,\cite{bai2000templates,cullum2002lanczos}
we have been careful not to overconverge the Lanczos iterative process
by not setting the Lanczos error to be extremely small.

This paper is organized as follows. Sec.~\ref{sec:method} presents the Krylov-space approach 
for the calculation of the correction-vector. Sec.~\ref{sec:results} 
applies our equations to the Heisenberg spin chain, 
to the t-J model, and to the Hubbard model.
We have calculated the $S(k,\omega)$ for the Heisenberg and Hubbard models, 
while the $A(k,\omega)$ for a t-J chain. Section~\ref{sec:tjladder} 
shows that Krylov approach can efficiently calculate the dynamical spin structure
factor of the t-J model on a ladder geometry.
Section~\ref{sec:performance} compares 
the frequency resolution and the computational performance 
of the conjugate gradient method with the Krylov-space approach  
for the simulated models. 
The last section presents a summary and conclusions.

\section{Conjugate Gradient and Krylov methods}\label{sec:method}

This section briefly recalls the basics of the correction-vector method. 
One is interested in the calculation of the Green's function 
\begin{equation}\label{eq:green}
G(z) = -\frac{1}{\pi}\langle\psi_{0}|\hat{B}\frac{1}{z+E_{0}-\hat{H}}
\hat{A}|\psi_{0}\rangle,
\end{equation}
where $|\psi_{0}\rangle$ is the ground state of some Hamiltonian $\hat{H}$ with 
ground-state energy $E_{0}$, 
$\hat{A}$ and $\hat{B}$ are operators associated with the dynamical correlation 
function to be calculated,
and $z\equiv\omega+i\eta$, where $\omega$ is the real frequency and $\eta$ is 
a positive constant that provides a finite broadening of the Green's function peaks. This procedure is made rigorous by realizing that
what we call Green's functions are actually distributions.

The correction-vector associated with Eq.~(\ref{eq:green}) is defined by
\begin{equation}\label{eq:cv}
|x(z)\rangle = \frac{1}{z+E_{0}-\hat{H}}|A\rangle,
\end{equation}
where the vector $|A\rangle\equiv \hat{A}|\psi_{0}\rangle$ is assumed to be real. 
Within the traditional 
formulation of DMRG,\cite{re:Kuhner1999}
the correction-vector method uses a multi-target 
approach: at each step of the DMRG algorithm, 
one targets the ground state of the system $|\psi_{0}\rangle$, 
the vector $|A\rangle$ and the $|x(z)\rangle$ in the reduced density matrix,
for each frequency value $\omega$ and broadening $\eta$. 
After one obtains the correction vector, one can calculate
the Green's function $G(z)$ using 
\begin{equation}
G(z) = -\frac{1}{\pi}\langle\psi_{0}|\hat{B}|x(z)\rangle.
\end{equation}

A direct approach to calculate the correction-vector is to solve 
the following set of linear equations in the local DMRG basis:
\begin{equation}\label{eq:GMRES}
(z+E_{0}-\hat{H})|x(z)\rangle = |A\rangle.
\end{equation}
Eq.~(\ref{eq:GMRES}) could be solved with the 
generalized minimal residual method (GMRES), which uses 
the Arnoldi algorithm to find a generalized decomposition of the matrix $M\equiv z+E_{0}-\hat{H}$ onto 
a Krylov subspace of much smaller dimension than the local Hilbert space.\cite{saad2003iterative} 
But such approach would have the drawback that the matrix $M\equiv z+E_{0}-\hat{H}$ is not 
Hermitian due to the presence of the factor $\eta>0$ in $z=\omega+i\eta$. 
The convergence error of the algorithm is given by the condition number of 
the matrix $M$, $\kappa(M)$.\cite{saad2003iterative} 
The larger the condition number, the greater is the number of iterations needed for 
solving the set of equations, and the smaller is the improvement of the solution at each iteration step. 

To avoid dealing with non Hermitian matrices, the imaginary part 
of the correction-vector, $|x(z)\rangle_{\text{Im}}$, is calculated 
by solving the following set of linear equations
\begin{equation}\label{eq:seteq}
[(E_{0}+\omega-\hat{H})^{2}+\eta^{2}]|x(z)\rangle_{\text{Im}}=-\eta|A\rangle,
\end{equation}
instead of Eq.~(\ref{eq:GMRES}). The real part of the 
correction-vector $|x(z)\rangle_{\text{Re}}$ is then calculated from
its imaginary part using 
\begin{equation}
|x(z)\rangle_{\text{Re}}=\frac{\hat{H}-E_{0}-\omega}{\eta}|x(z)\rangle_{\text{Im}}.
\end{equation}
In this case, the matrix $M'\equiv[(E_{0}+\omega-\hat{H})^{2}+\eta^{2}]$ is real, symmetric, and 
positive definite, 
therefore at each DMRG step, Eq.~(\ref{eq:seteq}) can 
be solved with the conjugate gradient 
method.\cite{saad2003iterative} 
This method can be thought of as a particular case of GMRES for real symmetric and 
positive definite matrices.\cite{saad2003iterative}
The main problem with this second approach is 
that the condition number of the matrix $M'$, $\kappa(M')$ in Eq.~(\ref{eq:seteq}) 
is larger: roughly 
the \emph{square} of the condition number of Eq.~(\ref{eq:GMRES}). 

Recall that, for a system of inhomogeneous set of equations $M\textbf{x}=\textbf{b}$ 
for the unknown vector $\textbf{x}$,
the error in the conjugate-gradient algorithm at step $k$ is given by the 
norm of the residual $\textbf{r}_k=\textbf{b}-M\textbf{x}_k$, where $\textbf{x}_k$ 
is the approximate solution at that step \cite{saad2003iterative}.
In the next sections, we assume that a measure of the error
for the conjugate-gradient algorithm is given by the number of iterations required for 
reaching a certain tolerance on the solution.

In this paper, we propose an alternative for calculating 
the correction-vector in Eq.~(\ref{eq:cv}). At each step of the DMRG, we perform a Lanczos 
tri-diagonalization of the Hamiltonian and a successive diagonalization  
in the current DMRG basis; 
the vector $|x(z)\rangle$ is then calculated \emph{directly} as 
\begin{equation}\label{eq:cvapprox}
|x(z)\rangle = V^{\dag}S^{\dag}\frac{1}{E_{0}+\omega-D+i\eta}SV|A\rangle,
\end{equation}
where $D$ is the diagonal form of the Hamiltonian operator $\hat{H}$ in the current basis.
We have assumed that
\begin{equation}\label{eq:Happrox}
\hat{H} = V^{\dag}TV = V^{\dag}S^{\dag}DSV
\end{equation}
is a faithful representation of the Hamiltonian
when applied to the starting vector. $V$ represents the matrix of the Lanczos vectors spanning the Krylov space, and $T$ 
the representation of the Hamiltonian in tridiagonal form. 
The tri-diagonalization is then followed by a small full 
diagonalization of $T$, yielding the matrix of eigenvectors $S$.
The first equality in Eq.~(\ref{eq:Happrox}) 
holds with the Lanczos error, unlike in the conjugate gradient method,
where the error is separate from the DMRG. Therefore, for the Krylov method 
the error coincides with the Lanczos error which 
occurs in the tri-diagonalization of the Hamiltonian $\hat{H}$.
The accuracy of the approximation Eq.~(\ref{eq:cvapprox}) can be estimated from the high frequency
expansion of the Green's function Eq.~(\ref{eq:green}) in the limit of $\eta\rightarrow0$\cite{re:Balzer2012}
\begin{equation}
\lim_{\eta\rightarrow0}G(z)=\sum_{r=0}^{\infty}\frac{1}{\omega^{r+1}}\langle\psi_{0}|\hat{B}(\hat{H}-E_{0})^{r}\hat{A}|\psi_{0}\rangle.
\end{equation}
Using approximation Eq.~(\ref{eq:Happrox}) in the above equation---usually satisfied after $n$ Lanczos iterations for a certain required accuracy---the
approximated Green's function reproduces the first $n$ coefficients of the expansion
\begin{equation}\label{eq:greenapprox}
\lim_{\eta\rightarrow0}G(z)\simeq\sum_{r=0}^{\infty}\frac{1}{\omega^{r+1}}\langle\psi_{0}|\hat{B}[V^{\dag}(\hat{H}-E_{0})V]^{r}\hat{A}|\psi_{0}\rangle,
\end{equation}
because $(\hat{H}-E_{0})^{r}\hat{A}|\psi_{0}\rangle\simeq[V^{\dag}(\hat{H}-E_{0})V]^{r}\hat{A}|\psi_{0}\rangle$ within Lanczos error because the vector 
belongs to the Krylov space for 
all $r\leq n-1$. We have here neglected the error in the determination of the ground state $|\psi_{0}\rangle$.
The expansion Eq.~(\ref{eq:greenapprox}) therefore produces a good approximation 
for the first $n$ moments of the spectral function $\int d\omega\, \omega^{r}A(\omega)$, with $r\leq n-1$,
where $A(\omega)=-\text{Im}[G(z)]/{\pi}$.\cite{re:Balzer2012}
As will be confirmed numerically in the next sections, the accuracy and efficiency of this approach is then
 best at low frequencies, and worsens for high excitation energies. In our later analysis, we shall find that the number 
of iterations required to reach a certain accuracy in our Krylov approach 
is of the order of $\sqrt{\kappa(M')}$, showing evidence that that the conjugate gradient method 
is less efficient.
When the number of Lanczos iterations is sufficiently small, then the dimension of 
the matrix $T$ is of the order of a few hundreds; for all the models and geometries investigated
in this paper, we find that about $150$ iterations $n$ 
are at most needed in our simulations. The computational cost of the diagonalization 
$S$ is then negligible.
On the contrary, the number of conjugate gradient iterations necessary to obtain 
the same accuracy is much larger. 

Our method is similar to DMRG continued-fraction-expansion or CFE in that 
it uses a tridiagonal decomposition of the Hamiltonian to compute the correction 
vector. The approach is therefore equivalent to 
fully diagonalizing the tridiagonal 
Hamiltonian and obtaining the matrix $S$ on the one hand, or doing a CFE on the other.
Our method is \emph{dissimilar} to DMRG CFE in that it is a correction-vector method, and thus Lanczos vectors are not 
targeted. In fact, DMRG CFE refers to the method where all or some of the Lanczos vectors are targeted in the reduced
density matrix at each DMRG step. In the original paper\cite{re:Hallberg1995} by 
K. Hallberg all the Lanczos vectors are targeted,
while in an improved version\cite{re:Dargel2011}
 by P.E. Dargel \emph{et al.} only three Lanczos
vectors at a time are targeted as the lattice is swept.

\section{Numerical results}\label{sec:results}

\subsection{Dynamical spin structure factor of Heisenberg chains}\label{sec:res1}

We begin by studying the antiferromagentic Heisenberg model on an open 1D chain of $L$ sites
\begin{equation}\label{eq:Heis}
\hat{H}_{Heis}= J \sum_{i=1}^{L-1} \textbf{S}_{i} \cdot \textbf{S}_{i+1},
\end{equation}
where $\textbf{S}_{i}$ denotes the spin operator at site $i$. We choose $J=1$ as unit of energy 
for this model. As in the original paper of K\"{u}hner and 
White,\cite{re:Kuhner1999}
the longitudinal dynamical spin structure factor is calculated with correction-vector DMRG as
\begin{equation}\label{eq:Sijw}
S_{j,c}(\omega+i\eta)=-\frac{1}{\pi}\text{Im}\Bigg[\langle \Psi_{0}|S^{z}_{j}
\frac{1}{\omega-\hat{H}+E_{0}+i\eta}S^{z}_{c}|\Psi_{0}\rangle\Bigg],
\end{equation}
where the operator $S^{z}_{c}$ is applied to the ground state at the center of the 
lattice for calculating the correction vector $|x(\omega+i\eta)\rangle$. 
Then, the correlator 
\begin{equation}\label{eq:Sijw2}
S_{j,c}(\omega+i\eta)=-\frac{1}{\pi}\text{Im}\Bigg[\langle \Psi_{0}|S^{z}_{j}|x(\omega+i\eta)\rangle\Bigg]
\end{equation}
is computed for all the sites $j$ of the lattice, and for a fixed value of frequency 
$\omega$ and broadening $\eta$. The above quantity is finally transformed to momentum space as 
\begin{equation}\label{eq:Skw}
S(k,\omega)=\sqrt{\frac{2}{L+1}}\sum_{j=1}^{L-1}\sin((j-c)k)S_{j,c}(\omega+i\eta),
\end{equation}
where the quasi-momenta $k=\frac{\pi n}{L+1}$ with $n=1,..,L$ are appropriate for 
open boundary conditions.
The DMRG implementation used throughout this paper is discussed in 
the supplemental material, which can be found 
at~\url{https://drive.google.com/open?id=0B4WrP8cGc5JHX3h6M25zUVhiQmc}.

\begin{figure}
\centering
\includegraphics[width=8.5cm]{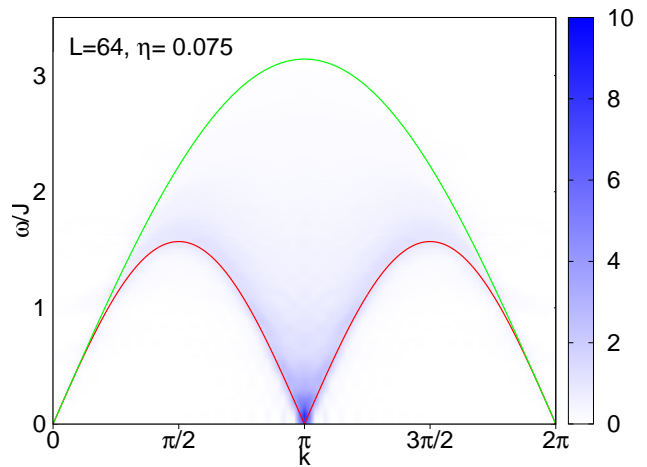}
\caption{(Color online) $S(k,\omega)$ for an antiferromagnetic Heisenberg chain of $L=64$ 
sites calculated with the Krylov-space approach. We have used $\eta=0.075$, $m_{min}=64$ and $m=1000$ 
DMRG states, with a truncation error kept at $10^{-8}$. Solid (red) lines indicate the 
lower boundary of the Bethe ansatz spin excitation continuum, $\omega^{l}(k)$; see Eq.~(\ref{eq:dCP}). 
Dashed-dotted (green) line indicates the upper boundary of the spin excitation continuum,
$\omega^{u}(k)$.} \label{fig1}
\end{figure}

It is known from the Bethe ansatz
solution\cite{re:Cloizeaux1962,re:Muller1979,re:Muller1997,
re:Caux2006} 
of Hamiltonian Eq.~(\ref{eq:Heis}) that the 
upper and lower boundary of the spin excitation manifold of the infinite system, 
called the des Cloiseaux-Pearson (dCP) dispersion relations, are given by
\begin{equation}\label{eq:dCP}
\omega^{l}(q)=(J\pi/2)\sin(q),\;\;\omega^{u}(q)=J\pi|\sin(q/2)|,
\end{equation}
where $q$ represents the momentum and not the quasi-momentum. Moreover, it is known that 
the $S(k,\omega)$ diverges as
\begin{equation}\label{eq:diverg}
\begin{aligned}
S(k,\omega) &\sim [\omega-\omega^{l}]^{-1/2}\sqrt{\ln[1/(\omega-\omega^{l})]}~\text{for}~k\neq\pi,\\
S(\pi,\omega) &\sim \omega^{-1}\sqrt{\ln(1/\omega)},
\end{aligned}
\end{equation}
as $\omega$ approaches the lower boundary $\omega^{l}$ from above. This divergence has its profound
origin in the Luttinger liquid nature of the ground state, and describes
the instability of the model toward antiferromagnetic ordering. Numerically, 
because one has always
finite size systems, one usually cuts off the divergences at $\omega-\omega^{l}\simeq 1/L$, so that 
one has peaks of finite height
\begin{equation}\label{eq:max}
\begin{aligned}
\text{max}[S(k,\omega)] &\sim [L\ln(L)]^{1/2}~\text{for}~k\neq\pi,\\
\text{max}[S(\pi,\omega)] &\sim L\ln(L)^{1/2}.
\end{aligned}
\end{equation} 
Fig.~\ref{fig1} shows the dynamical 
spin structure factor of an antiferromagetic ($J=1$) Heisenberg spin chain of size $L=64$.
The Krylov-space correction-vector method described in the previous section has been applied.
We use $m_{min}=64$ and a maximum of $m=1000$ DMRG states by keeping the truncation error no bigger than $10^{-8}$. Moreover, we set the maximum number of Lanczos iterations necessary to 
calculate the correction-vector to $1000$, keeping the tri-diagonalization 
error no bigger than $10^{-7}$. 
For all the models and geometries investigated in this paper, we have found that the number of Lanczos iterations needed is at most about $150$.
Even at a finite system size, the dCP relations, which are 
exact in thermodynamic limit, are very well reproduced by the DMRG data.

\begin{figure}
\centering
\includegraphics[width=8.5cm]{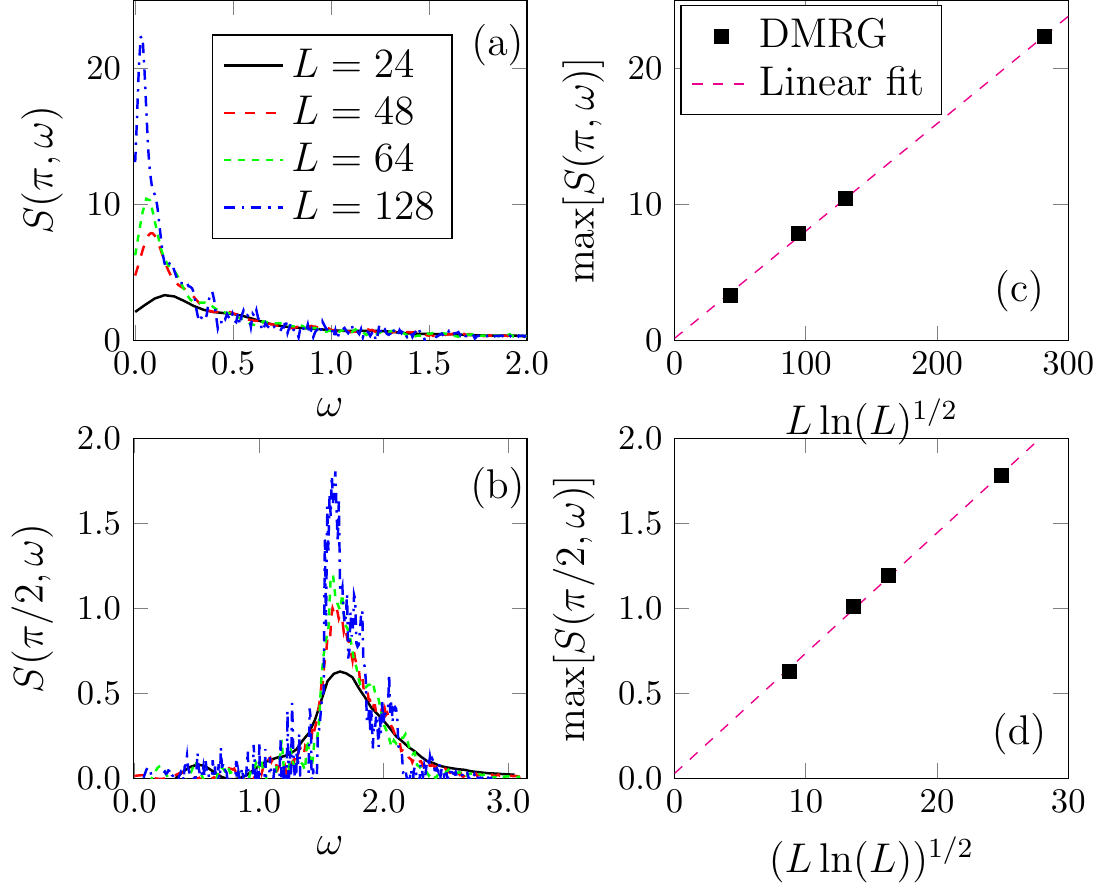}
\caption{(Color online) Panel (a) (panel (b)): $S(\pi,\omega)$ ($S(\pi/2,\omega)$) for a Heisenberg 
chain for different system sizes. Panel(c): max[$S(\pi,\omega)$] extracted from panel (a) as a 
function of $L\ln(L)^{1/2}$. Panel (d): max[$S(\pi/2,\omega)$] extracted from panel (b) as a 
function of $(L\ln(L))^{1/2}$.} \label{fig2}
\end{figure}
As explained thoroughly in ref.~\onlinecite{re:Jeckelmann2002}, finite size scaling of 
dynamical correlation functions with correction-vector DMRG should be done carefully: 
the artificial broadening $\eta$ of the spectral function 
peaks should be rescaled as function 
of system size, such that $\text{lim}_{L\rightarrow\infty}\eta(L)=0$. 
We find that the scaling of the peaks maxima in Eq.~(\ref{eq:max})
is well reproduced by imposing $\eta=c/L$, where $c$ is a constant of the order of 
the width of the full spectrum. Panels (a) and (b) of fig.~\ref{fig2} show the cuts of the
$S(k,\omega)$ spectrum at $k=\pi$ and $k=\pi/2$ as a function of $\omega$ for 
different system sizes. When using open boundary conditions,
we use the convention that $k=\pi$ refers to $k={\pi L}/(L+1)$, 
while $k=\pi/2$ refers to $k={\pi L}/(2(L+1))$. We have also verified that, for all 
system sizes investigated, there is no qualitative difference between open and periodic 
$k$ values used in the Fourier transform. 
Panel (a) shows a shift of the peak position toward $\omega\rightarrow0$, which is the 
expected position of the divergence at $k=\pi$ in the thermodynamic limit. The position of the 
peak in panel (b) is approximately constant as a function of the system size 
and approximately close to the expected thermodynamic limit value $\omega^{l}(\pi/2)\simeq\pi/2$. 
Panels (c) and (d) show the maxima of the peaks obtained in panel (a) and (b) as a function 
of the system size. With a linear fit, we have verified that the peaks 
maxima have the correct scaling as described by the relations in Eq.~(\ref{eq:max}). 

Fig.~\ref{fig3}, panels (a) and (c) show the dynamical structure factor for a 
Heisenberg chain with $L=48$ sites, calculated with the Krylov-space and the conjugate gradient 
method, respectively. In this figure, $\eta=0.1$ and $m=800$ DMRG states were kept 
in our simulations.
The frequency resolution of the Krylov-space approach is much better than 
that provided by the conjugate gradient method. 
While in the Krylov-space approach the spectral weight outside the region defined 
by the dCP relations is practically zero, the conjugate gradient 
method gives a spectral weight spread everywhere in the frequency interval investigated, 
with much less defined peak features.
\emph{We have tried to use the
same parameters for both approaches, but could not do so precisely, as we now explain.}

We have set the maximum number of 
conjugate gradient iterations to $1000$ 
trying to keep the error no bigger than $10^{-7}$. 
Unfortunately, we have found that, for most of the frequency 
interval investigated, a number of iterations much larger than $1000$ is necessary to 
reach convergence. \emph{The conjugate-gradient error can thus be as high as $10^{-1}$.} 
This explains why the spectrum shown in panel (c) has much less 
frequency resolution than the spectrum calculated with the Krylov-space approach.
Panel (e)
shows a more detailed comparison between the spectra calculated with the two approaches. 
Cuts at $k=\pi$ and $k=\pi/2$ are examined, indicating that conjugate-gradient is able 
to capture only qualitatively the main features of the spectra.
For the cut at $k=\pi$ the position of the peak at low frequency $\omega\simeq0$ is
correctly reproduced, but the spectral weight is smaller than that provided by the 
Krylov-space approach by a factor of $3$. 
A different behavior is observed for the peak at $k=\pi/2$. 
Here, the position of the spectral peak is shifted to lower frequency, 
while the spectral weights are also redistributed 
to lower frequencies; see dashed (red) line in panel (e).

\subsection{Dynamical spin structure factor of Hubbard chains}\label{sec:resHubbard}

\begin{figure}
\centering{\includegraphics[width=8.5cm]{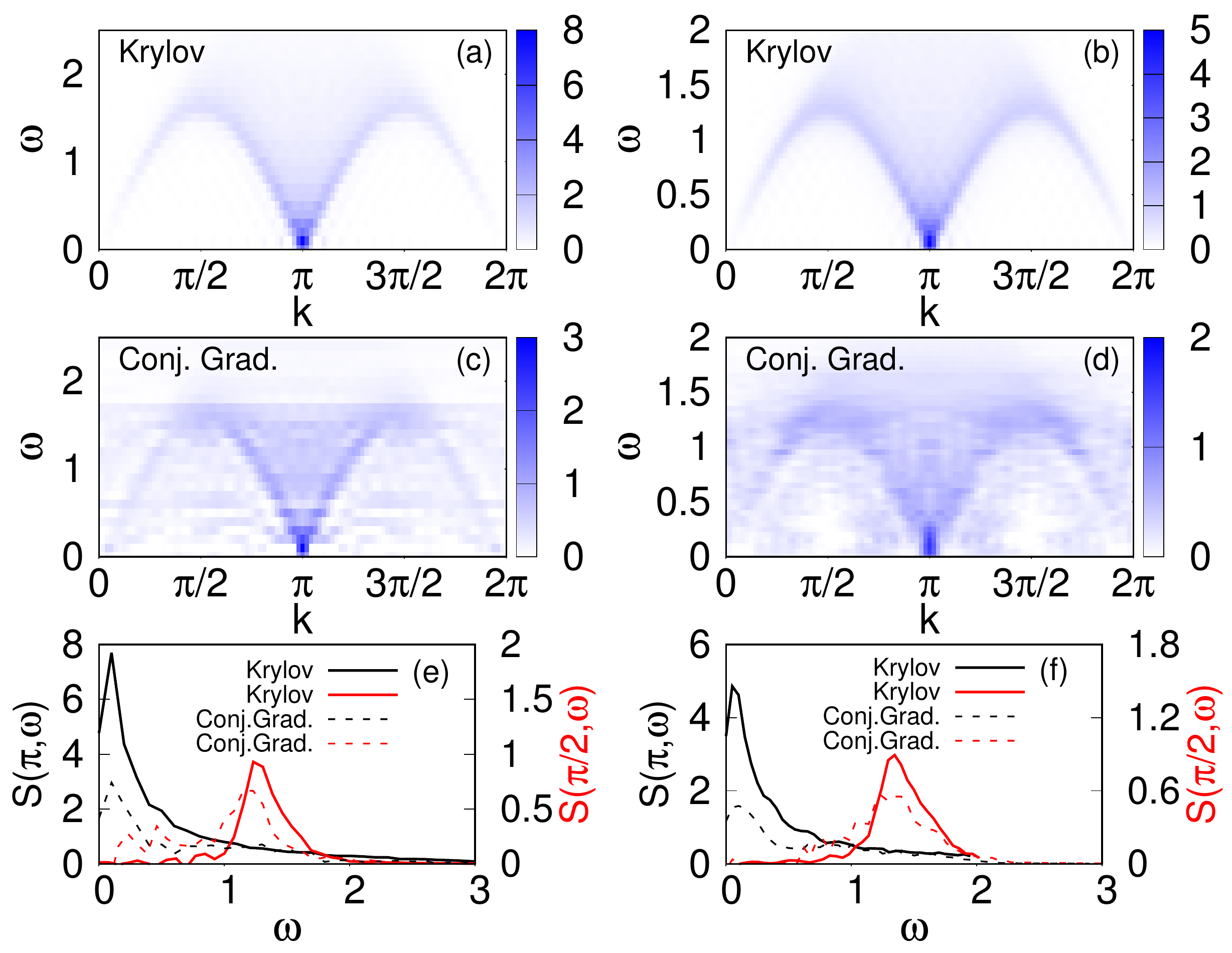}}
\caption{(Color online) Panels (a-b-c-d): $S(q,\omega)$ calculated with the Krylov-space
approach (a-b), and the conjugate gradient method (c-d) for a Heisenberg 
(panels (a-c)) and a Hubbard 
(panels (b-d)) chain of $L=48$ sites, using $\eta=0.1$ and $m=800$ DMRG states
($m_{min}=64$). Panel (e-f) shows cuts at $k=\pi$ and $k=\pi/2$ of the 
spectra in (a-c) and (b-d) as a function of frequency. In the conjugate gradient 
method, a maximum of $1000$ iteration steps 
has been imposed, regardless of the error.} \label{fig3}
\end{figure}

In the same figure~\ref{fig3}, panels (b) and (d) show the dynamical spin structure 
factor calculated with the two methods outlined above for a Hubbard chain of $L=48$ sites 
at half-filling and at $U/t=4$. The Hubbard model Hamiltonian is given by
\begin{equation}\label{eq:Hubb}
\hat{H}_{Hubbard}= -t\sum_{i=1}^{L-1} (c^{\dag}_{i} c_{i+1}+\text{h.c.})
+U\sum_{i=1}^{L}n_{\uparrow,i}n_{\downarrow,i},
\end{equation}
where we have used standard notation. In these panels, $t=1$ is assumed as unit of energy.
Notice the similarity between the Hubbard and the Heisenberg chain spectrum in panels (a) and (b)
calculated with Krylov-space approach.
The Bethe ansatz solution\cite{re:Ogata1990} at half-filling 
and in the limit of 
large Coulomb repulsion $U\gg t$, confirms that the Hubbard model on a chain is gapless in the spin sector. In this case, the Hubbard model behaves as the Heisenberg model.
Similar to the Heisenberg case, the quality of the spectrum obtained with the conjugate gradient
is much worse than that obtained with Krylov. Here again, the spectral weight is distributed
everywhere outside the region delimited by the dCP relations. 
A detailed comparison of cuts of the spectrum at $k=\pi$ and $k=\pi/2$ in panel (f) show 
again that the position of the spectral peaks is only qualitatively captured 
by the conjugate gradient.
   
\subsection{Spectral function of a t-J chain}\label{sec:res2}

\begin{figure}
\centering
\includegraphics[width=9.5cm]{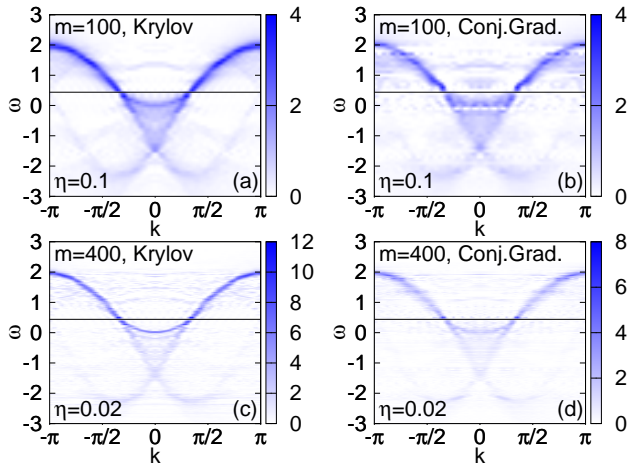}
\caption{(Color online) Panel(a-c): $A(k,\omega)$ for a t-J chain of $L=48$ sites
with Krylov-space approach for $m=100$ and $\eta=0.1$ (a), $m=400$ $\eta=0.02$ (c). Panels (b-d): same as
panels (a) and (c) but with the conjugate gradient method.
The chemical potential $\mu\simeq0.44$ is indicated with a solid black line.} \label{fig4}
\end{figure}

Fig.~\ref{fig4} shows the spectral function $A(k,\omega)$ for a one dimensional t-J chain 
of $L=48$ sites and filling $N/L=2/3$. The t-J model Hamiltonian is given by
\begin{equation}\label{eq:tJ}
\hat{H}_{tJ}= -t\sum_{i=1}^{L-1} (c^{\dag}_{i} c_{i+1}+\text{h.c.})
+J \sum_{i=1}^{L}\big(\textbf{S}_{i}\cdot\textbf{S}_{i+1}-n_{i}n_{i+1}/4\big),
\end{equation}
where we have used standard notation. For this model, we choose the unit of energy $t=1$, and study 
the case where $J=0.5$. The spectral function $A(k,\omega)=\sum_{\xi=\pm}A^{\xi}(k,\omega)$
consists of two branches: the photoemission $\xi=-$, and the anti-photoemission $\xi=+$.
\begin{equation}\label{eq:Aijw}
A^{\xi}_{j,c}(\omega+i\eta)=-\frac{1}{\pi}\text{Im}\Bigg[\langle \Psi_{0}|c^{\bar{\xi}}_{j}
\frac{1}{\xi\omega-\hat{H}+E_{0}+i\eta}c^{\xi}_{c}|\Psi_{0}\rangle\Bigg],
\end{equation}
where $\bar{\xi}=-\xi$ and $c^{-}_{x}\equiv c_{x}$ (destruction operator), 
while $c^{+}_{x}\equiv c^{\dag}_{x}$ (construction operator) at site $x$. 
The two branches of the spectrum are calculated with correction-vector DMRG and 
Fourier transformed to momentum space, as described in the previous subsection 
for the dynamical spin structure factor $S(k,\omega)$.

Panels (a) and (b) of Fig.~\ref{fig4} show the spectral function of the t-J chain calculated with the 
Krylov-space 
and conjugate gradient approaches with $\eta=0.1$ and keeping a small number 
$m_{min}=m=100$ of DMRG states. The spectral function of the t-J model 
contains the phenomenon of spin-charge separation. 
Below the Fermi level (indicated by a solid black line), the spectral 
weight is concentrated on the spinon and holon bands. The spinon band forms an arch of 
amplitude $J\simeq0.5$ connecting the two Fermi points $(-k_F,\omega-\mu=0)$ 
and $(k_F,\omega-\mu=0)$. Because the filling is $N/L=2/3$, 
then $k_F=\pi/3$. The holon bands depart from the spinon band at the Fermi points 
in two approximately straight lines with slope $2t/k_{F}$.
As expected for a Luttinger liquid, the shadow bands extend in frequency well beyond $|k_{F}|$.

Panel (c) and (d) of Fig.~\ref{fig4} show the spectral function with a larger number of DMRG 
states, $m=400$. Regardless of the method used, the frequency resolution of the 
spectrum is improved. The frequency 
resolution of the holon and spinon branches of the spectrum is worse in the conjugate 
gradient case. Nevertheless, the qualitative features of the spectrum are captured 
by the conjugate gradient method too.

\subsection{Dynamical spin structure factor of a t-J ladder}\label{sec:tjladder}

We now show that our Krylov based approach can be used to calculate efficiently 
the magnetic excitation spectrum of systems with more complex geometries than chains. As a case study, we analyze a t-J model on a ladder geometry, with Hamiltonian
\begin{align}\label{eq:HtJladder}
H&=-t_{x} \sum\limits_{\substack{\langle i,j\rangle\\ \sigma,\gamma=0,1}}(c^\dagger_{i,\gamma,\sigma}
c_{j,\gamma,\sigma} + h.c.) - t_y \sum\limits_{i,\sigma} c^{\dag}_{i,0,\sigma}
c_{i,1,\sigma} \nonumber\\
&+ J \sum\limits_{i,\gamma=0,1} \Big(\vec{S}_{i,\gamma}\cdot\vec{S}_{i+1,\gamma}-\frac{1}{4} n_{i,\gamma}
n_{i+1,\gamma}\Big).
\end{align}
Recently, 
the ground state properties and the spectral function $A(k,\omega)$ in the limit 
of one hole doping have been studied with DMRG on large system sizes.\cite{re:White2015,re:Zhu2015}
Yet it is well known that finite dopings are more difficult to treat with DMRG, 
therefore our calculation required a comparable effort.  
In the past, this model was thoroughly studied in the context of cuprates.\cite{re:Dagotto1992,re:Dagotto1996,re:Dagotto1999}
In the undoped limit, it has been well established that 
the t-J model has a spin gap due to the particular 
ladder geometry, and, as in the chains' case, the physics can be described in terms of the 
Heisenberg ladder. Upon doping,
superconductivity mainly occurs in the d-channel, as described in refs.~\onlinecite{re:Dagotto1992,re:Tsunetsugu1994,re:Maier2008,re:Poilblanc2003,re:Hayward1995}. 
As suggested in 
ref.~\onlinecite{re:Scalapino1998}, neutron scattering data could provide important evidence for 
a pairing mechanism based on the exchange interaction $J$. 
The physics 
of t-J (and Hubbard) two leg ladders has been studied with many techniques ranging 
from bosonization to DMRG to exact diagonalization. The mostly studied case has been the isotropic 
regime, where $t_x=t_y=t$. The possibility of pairing has been also 
investigated in the anisotropic limit, where $J_y/J_x\neq1$ and $t_y/t_x\neq1$. At half-filling
a Heisenberg description applies, and the spin gap has been shown to persist until very small 
values of the $J_y$ parameter.\cite{re:Dagotto1992}
Away from half-filling, the authors of ref.~\onlinecite{re:Riera1999}
studied the spin gap and the superconducting binding energy of the hole pairs showing that
they can be maximized by tuning the anisotropic ratios to $t_y/t_x \simeq 1.25$, and 
$J_y/J_x \simeq 1.56$.

\begin{figure}
\centering{\includegraphics[width=8.5cm]{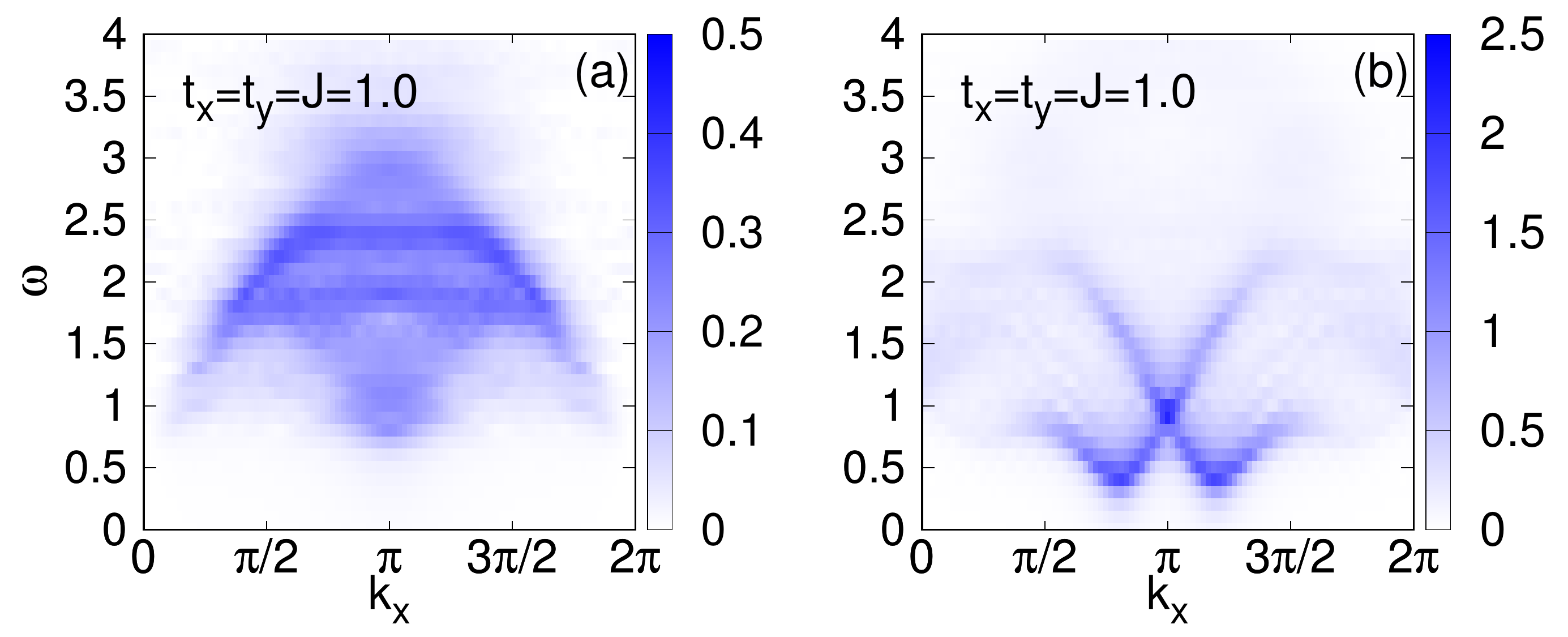}}
\caption{(Color online) Panel (a): $S(k_x,0,\omega)$ component for two leg ladder t-J model 
with $L=48\times 2$ sites at $t_x=t_y=J=1.0$ and filling $n=0.8125$. 
$m=600$ states are kept in the DMRG simulations. 
Panel (b): $S(k_x,\pi,\omega)$ component of the spectrum for the same parameter values 
of panel (a).} \label{fig:tJladder}
\end{figure}

Fig.~\ref{fig:tJladder} shows our results obtained with DMRG for the spin structure factor. 
In the case of the ladder,  
the dynamical structure
factor in momentum space has two components
because the momentum in the y direction has only two possible values: $k_y=0$ and $k_y=\pi$. 

Let $t_x=1$ be our unit of energy, and let us study the spectrum 
for the set of parameters $t_y=J_y=J_x=1$. 
The characteristics of the spectrum are 
completely different from the case of decoupled chains (not shown). 
Indeed, a spectral gap appears clearly, both in the 
$k_y=0$ and in the $k_y=\pi$ components of the spectrum. 
In particular, 
a very dispersive gap arises in the $k_y=0$ component with amplitude at $k=0$  
$\omega_{gap}(0)\simeq0.75$. In the same
component, the spin excitations form a triangular structure where all 
the spectral weight is concentrated. Two separate dispersive arcs can be noticed.
The $S(k,\pi,\omega)$ component shows a very defined and peculiar shape. 
A spin gap can be observed, with $\omega_{gap}\simeq0.5$, a value smaller than the gap 
in the other component of the spectrum. 
Above that gap, the dispersion curves 
of the spin excitations describe the low energy boundary of the non-interacting tight-binding 
chain results. As in the case of a t-J chain, the spectral weight 
is redistributed to low energy, but the weight decreases rapidly away from the low boundaries.

\section{Computational performance}\label{sec:performance}

\begin{figure}
\centering{\includegraphics[width=8cm]{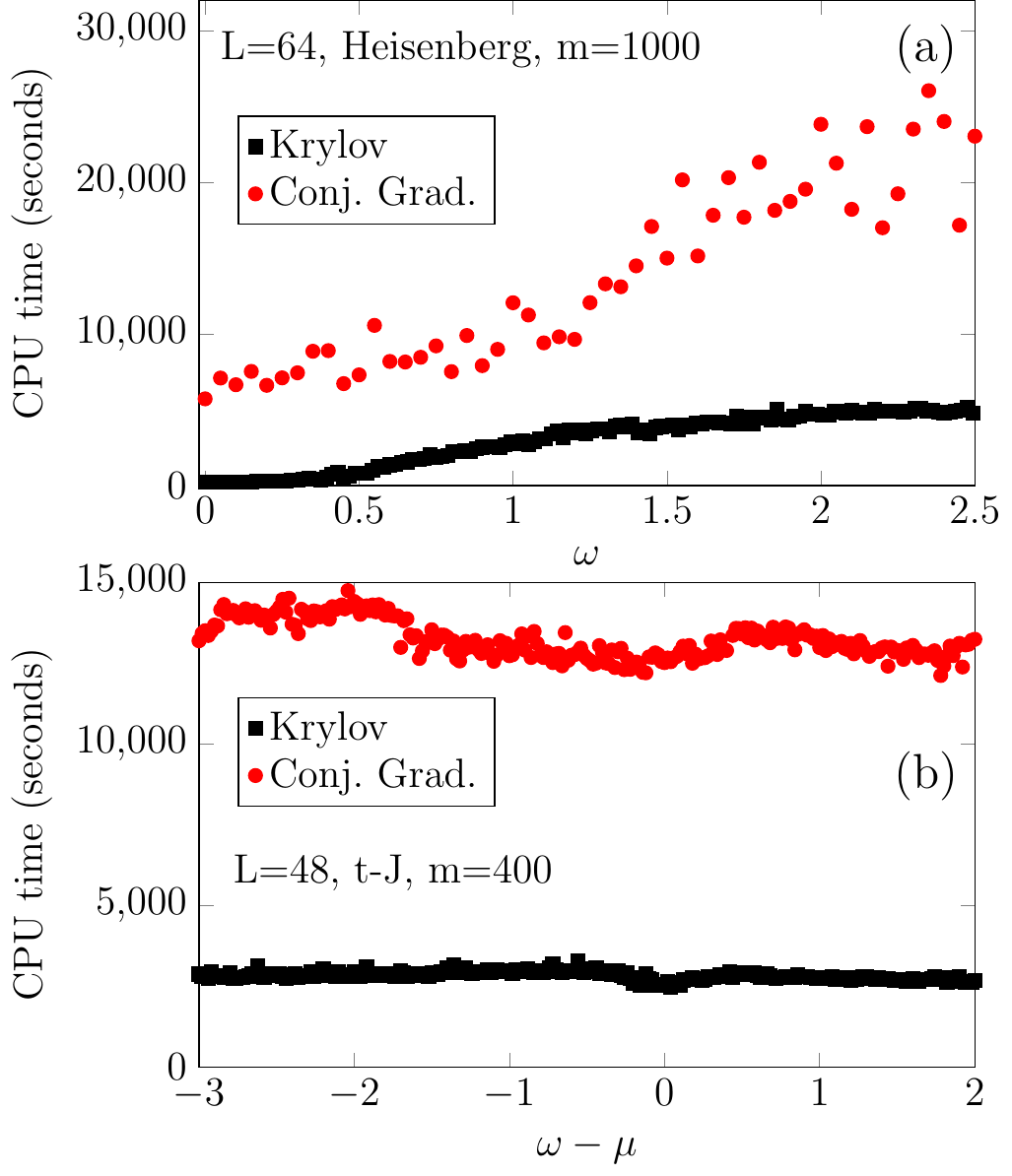}}
\caption{(Color online) Panel~(a): CPU times for a DMRG run as a function of the frequency 
$\omega$ performed with Krylov-space and Conjugate gradient methods for a Heisenberg model. 
Panel~(b): same as panel (a) but for the t-J chain model investigated in 
sec.~\ref{sec:res2}.} \label{fig5}
\end{figure}

This subsection compares the computational performance of the Krylov-space
and conjugate gradient methods for the dynamical spectra studied in the previous sections.
Panel (a) of fig.~\ref{fig5} shows the CPU time needed at each frequency $\omega$ 
for the two methods, when calculating the dynamical spin structure factor 
of the Heisenberg model, investigated in section~\ref{sec:res1}.
\begin{figure}
\centering{\includegraphics[width=8cm]{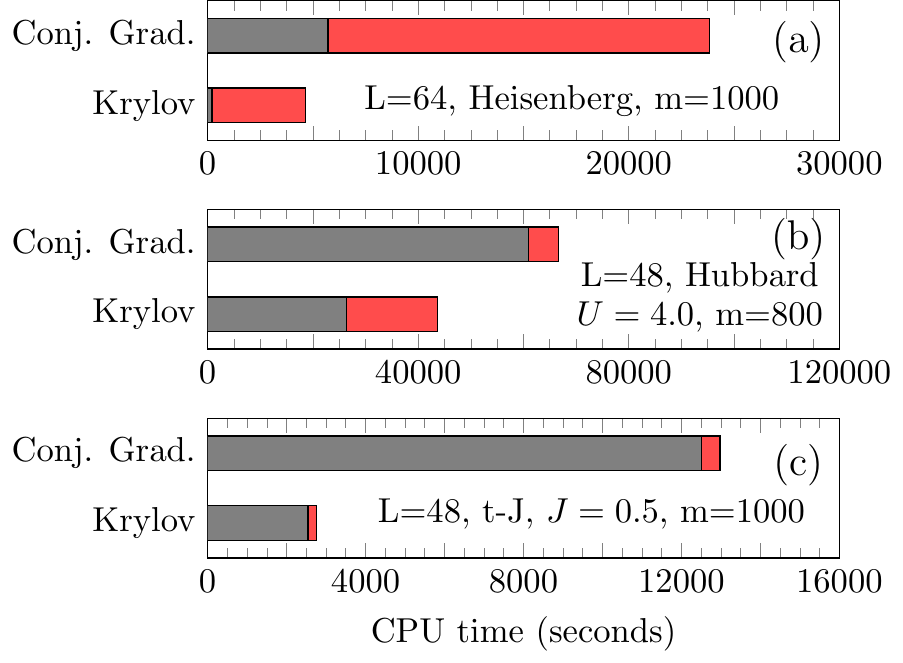}}
\caption{(Color online) CPU times for a DMRG run performed with Krylov and Conjugate 
gradient methods for a Heisenberg (panel (a)), Hubbard (panel (b)) and t-J model (panel (c)). 
In panels (a) and (b), the grey and red bars indicate the CPU times for $\omega=0$ and $\omega=2$, 
respectively. In panel (c), the grey and red bars indicate the CPU times for $\omega-\mu=0$ and 
$\omega-\mu=1$, respectively.} \label{fig6}
\end{figure} 
With the Krylov-space approach, the CPU time is smaller in the low frequency regime
than in the high frequency one. At larger frequencies, larger CPU times are needed.
As outlined in the previous section, we have set a maximum number of Lanczos iterations equal to $1000$,
and kept the error no larger than $10^{-7}$. This accuracy is reached in about 
$50$ iterations in the low frequency regime $0<\omega<0.5$, while the number of iterations needed
increases up to $150$ in the large frequency regime. 
The conjugate gradient method has a similar CPU time profile; see the circle (red) symbols. 
As already mentioned in the previous section, we have set the maximum number of conjugate gradient 
iterations to $1000$, and have considered the method converged if the error is smaller than $10^{-7}$. 
\emph{However, for most 
of the frequency interval investigated, the error can be as high as $10^{-1}$
with $1000$ iterations or less.
We have found that increasing the maximum number of 
iterations needed by the algorithm to reach an error less than $10^{-7}$ increases the CPU time even more that what we have shown in this paper}. As seen in 
 panel (a) of fig.~\ref{fig6}, the Krylov method 
is faster than the conjugate gradient by more the one order of magnitude
in the very low frequency regime. But it is not just faster, the conjugate gradient has not
even converged despite having used a large amount of CPU time and iterations. Panel (b) of fig.~\ref{fig6} compares
the CPU time performance of the two methods for the spectrum of the 
Hubbard model. Here the performance of the conjugate gradient is better, 
but still at least a factor of 
$3$ smaller than that of the Krylov-space approach.

Panel (b) of fig.~\ref{fig5} shows the CPU time as function of the frequency
for the spectral function $A(k,\omega)$ of the t-J model studied in sec.~\ref{sec:res2}.
The Krylov-space approach is again substantially faster than the conjugate
gradient.
Even if slightly visible, the Krylov approach has its best performance at low energy, 
that is, for $\omega-\mu\simeq0$, where it is $20\%$ faster than at higher frequencies. 
On the contrary, because of the convergence issues, we have not been able to
obtain a well defined CPU dependency on frequency for the conjugate gradient method.  
Panel (c) of fig.~\ref{fig6} shows the comparison between 
the two methods in the low frequency regime. As with the other models investigated, 
the Krylov method is faster than the conjugate gradient by a factor of $3$.

\section{Summary and Conclusions}\label{sec:conclusions}

This paper proposes an alternative method for computing correction-vectors 
based on a Krylov-space decomposition. We have tested the quality of our approach 
by studying the dynamical spin structure factor of a  
Heisenberg and Hubbard chain, and the spectral function of a t-J chain. 
We have also shown that the method is general, and applicable without 
restriction or further approximations to both more complex 
models and geometries. Beyond chains, we have shown in section~\ref{sec:tjladder} 
 that our method can be successfully applied to the calculation of the
dynamical spin structure factor of a t-J model on a ladder. 
The supplemental
material, which can be found at~\url{https://drive.google.com/open?id=0B4WrP8cGc5JHX3h6M25zUVhiQmc},
provides a pointer to the full open source code, input decks 
and additional computational details.

In all cases investigated, we have found that the Krylov-space approach
not only provides frequency spectra with much higher frequency resolution, but that it also requires
much less CPU time than the  the conjugate gradient 
method. 
When DMRG ground state itself uses Krylov space, then the calculation of
the correction vector integrates better with DMRG. In those implementations that
use the Davidson method\cite{re:Davidson1975} for ground state, the Davidson method
could perhaps be used to calculate the correction vector, as we have used  Krylov space here.
On the other hand, Chebyshev or Fourier-based methods for the computation of
spectral functions with DMRG should be computationally less expensive than 
the correction-vector method. 
But correction vector should be better suited for
problems where a constant resolution is needed, and where Chebyshev or Fourier-based methods 
might have limited resolution in parts of the spectrum.
For example, in neutron scattering and photoemission spectra---properties much needed to understand quantum magnets, superconductors, and transition metal oxides---resolving
fine features of the low-frequency spectrum is of great
importance.
The correction-vector method can be very precise in the estimation of these small energy gaps, 
with the only limitation given by the finite broadening $\eta$.

Using DMRG to 
obtain dynamical functions is meaningful because
the only other unbiased method, quantum Monte Carlo, does not
work directly in real frequency.
In the future,
we plan to extend our present work on dynamical properties 
to finite 
temperature.\cite{PhysRevB.93.104411,PhysRevB.90.060406}

\begin{acknowledgments}
We would like to thank E. Jeckelmann, S. R. Manmana,
 I. P. McCulloch, and F. A. Wolf for their feedback and suggestions.
This work was conducted at the Center for Nanophase Materials Sciences, sponsored
by the Scientific User Facilities Division, Basic Energy Sciences, Department
of Energy (DOE), USA, under contract with UT-Battelle. We acknowledge support by the DOE early
career research program.
\end{acknowledgments}

\bibliography{biblio}

\end{document}